\documentclass[ams,preprint]{revtex4}
\usepackage{amsmath}
\usepackage{epsfig}

\newcommand{\be}{\begin{equation}}
\newcommand{\ee}{\end{equation}}

\newcommand{\ber}{\begin{eqnarray}}
\newcommand{\eer}{\end{eqnarray}}

\begin{document}
\title{Discontinuities in fourth sound  waves in  superfluid helium}
\author{N.I. Pushkina}
\affiliation{M.V. Lomonosov Moscow State University, Research Computing Center,\\  Vorobyovy Gory, 
Moscow 119991, Russia}
\email{ N.Pushkina@mererand.com}

\begin{abstract}
Formation of fourth-sound shock waves in narrow channels filled with superfluid helium is studied. Physical and mathematical conditions at the surface of discontinuity  are established. These conditions differ somewhat from those in case of first- and second-sound waves.  The velocity of discontinuity coincides with that of  fourth sound. The jumps of  temperature and the superfluid velocity are shown to be of the first order as to the pressure jumps.

\end{abstract}
\maketitle

\section{Introduction}
In a narrow channel filled with superfluid helium the normal component can be locked by viscous forces when the free path of excitations is comparable or 
exceeds the diameter of a channel. But sound waves can still propagate through the superfluid component, and such kind of oscillations is called 
fourth-sound waves \cite{A, K}.  
 At rather low temperatures $T\le1,2^\circ K$ (see Ref. \cite{D}) the fourth-sound attenuation is so small that it does not exceed the uncertainty in the measurement of this quantity.  In any case the dominant source of  fourth-sound attenuation is the dissipation on the walls of a channel due to slipping of the normal component along the walls if there is such slipping. In case the normal component is completely locked  one can say that  fourth sound is a wave without dissipation. In this paper the formation of shock waves due to propagation of finite-amplitude fourth sound is studied.

\section{Theory}
Discontinuities of finite-amplitude first- and second-sound waves, that is in case of a bulk superfluid, were studied previously by Khalatnikov, see Ref. \cite{K}. In that case,
 at the surface of discontinuity the following quantities should satisfy the conditions of continuity: the mass flux density, momentum flux density, the force acting on the unit mass of the superfluid part of helium and the energy flux density. For fourth-sound shock waves the conditions on the surface of discontinuity are somewhat different. First, we cannot use the condition of continuity for the momentum flux density, because it does not include the dominant viscous force that reduces the normal component velocity to zero. There is also a peculiarity concerning entropy, and we shall dwell on it below. At the same time the following three continuity conditions should hold just as in the case of first- and second-sound shock waves.

1. The mass flux density
\begin{equation}
[{\bf j}]=[\rho_s{\bf v}_s+\rho_n{\bf v}_n]=0, \label{1} 
\end{equation}                                                                                                                   
square brackets mean the difference of the values of  quantities on different sides of the surface of discontinuity, $\rho_s$,\, $\rho_n$  are the superfluid and normal densities of helium, $v_s$,\,\,$v_n$   are the velocities of superfluid and normal parts.

2. The force acting on the unit mass of the superfluid part of helium
                                                                     
\begin{equation}                                                                               
\left[\mu+\frac{v_s^2}{2}\right]=0,     \label{2} 
\end{equation}                                                                                                                                               
 $\mu$  is the chemical potential which depends on thermodynamic variables, say, the fluid pressure $p$ and temperature  $T$  and on the squared difference of the normal and superfluid velocities $({\bf v}_n-{\bf v}_s)^2$ (in our case on $v_s^2$). 
   
 3. The energy flux density \cite{L}
\begin{equation}                                                                                   
\left[{\bf j}\left(\mu+\frac{v_s^2}{2}\right)+T\rho\sigma{\bf v}_n+\rho_n{\bf v}_n({\bf v}_n\cdot({\bf v}_n-{\bf v}_s))\right]=0,  \label{3} 
\end{equation}       
Here $\rho,\,\, \sigma$ are the fluid density and the entropy per unit mass. They also depend on pressure $p$ and temperature $T$ and on the squared difference of the 
normal and superfluid velocities. One can see that the first term in (\ref{3}) satisfies continuity conditions due to the equalities (\ref{1}) and 
(\ref{2}), and the second and third terms in (3) don't contribute to the energy flux because the normal velocity is reduced to zero.  Thus this condition can be regarded as already satisfied.    

Now let's admit that on one side of the discontinuity surface there is an undisturbed fluid, where the velocities $v_s$ and $v_n$  are equal to zero, and all the quantities are equal to their equilibrium values. Hence in Eqs. (\ref{1})$-$(\ref{3}) the supefluid and normal velocities in the coordinate system moving with the shock front velocity $u$ are equal to $-u$ on the undisturbed side, while on the other side they are equal correspondingly to $v_s-u$  and to $-u$ . As a result the continuity conditions (\ref{1})$,\,\, $(\ref{2}) in the coordinate system moving with the discontinuity velocity become
\begin{eqnarray}
\rho_0u=\rho_s(u-v_s)+\rho_nu=\rho u-\rho_sv_s, \label{4} \\
\mu_0+\frac{u^2}{2}=\mu+\frac{1}{2}(u-v_s)^2.   \label{5}   
\end{eqnarray}
The index $0$ denotes equilibrium values on the undisturbed side while the values on the other side are written without any index.      
    
Now we shall take into account that, as it was noted above, if there is no slipping of the normal component along the walls of the channel,  fourth sound in superfluid helium can be regarded as being a non-non-dissipating wave. Fourth sound propagates in a superfluid part of helium, and since superfluid flow involves no entropy transfer, the entropy value remains the same on both sides of the surface of discontinuity,      
\begin{equation}
\rho_0\sigma_0=\rho\sigma  \label{6} 
\end{equation}

 As a result we have ultimately three conditions on the surface of discontinuity unlike the case of  first- and second-sound shock waves where there are four conditions. But actually we do not need the fourth equation because the number of variables has been reduced to three since $v_n=0$, and we have just three equations.  
    As it was noted above the density, entropy and chemical potential are the functions of pressure  $p$,  temperature  $T$ and of the squared difference of the normal and superfluid velocities $({\bf v}_n-{\bf v}_s)^2$,  see Ref. \cite{L} (in our case $({\bf v}_n-{\bf v}_s)^2$  equals $v_s^2$),

\[\rho(p,T,v_s)=\rho(p,T)+\frac{1}{2}\rho^2v_s^2\frac{\partial}{\partial p}\frac{\rho_n}{\rho},\] 
\[\mu(p,T,v_s)=\mu(p,T)-\frac{1}{2}\frac{\rho_n}{\rho}v_s^2, \]
\[\sigma(p,T,v_s)=\sigma(p,T)+\frac{1}{2}v_s^2\frac{\partial}{\partial T}\frac{\rho_n}{\rho}.\]  

   We shall expand the thermodynamic quantities as a power series of $\triangle p=p-p_0$  and of  $\triangle T=T-T_0$ ,  limiting ourselves  with the first approximation.  In superfluid helium the thermal expansion coefficient is known to be extremely small, and for this reason we can neglect the dependence of the fluid density on temperature and the dependence of  entropy per unit mass on pressure.  With this we obtain Eqs. (\ref{4})$-$(\ref{6}) in the form
\begin{eqnarray}
u\frac{\partial \rho}{\partial p}\triangle p-\rho_sv_s=0,  \label{7} \\
\frac{1}{\rho}\triangle p-\sigma\triangle T-uv_s=0, \label{8} \\
\sigma\frac{\partial \rho}{\partial p}\triangle p+\rho\frac{\partial \sigma}{\partial T}\triangle T=0. \label{9} 
\end{eqnarray}
 This system of equations yields the following expression for the velocity $u$ of the shock front in the first approximation:
\[u^2=\frac{\rho_s}{\rho}\frac{\partial p}{\partial \rho}+\frac{\rho_s}{\rho}\frac{\sigma^2}
{\partial\sigma/\partial T}.\]
 Since the velocity of  first sound equals $u_1^2=\frac{\partial p}{\partial \rho}$, and the velocity of  second sound is equal to $u_2^2=\frac{\rho_s}{\rho_n}\frac{\sigma^2}
{\partial\sigma/\partial T}$, the expression for the velocity of the fourth sound shock front can be rewritten as
\[u^2=\frac{\rho_s}{\rho}u_1^2+\frac{\rho_n}{\rho}u_2^2.\]

     This expression is just the velocity of  fourth sound in superfluid helium. Thus one may say that weak jumps of pressure and temperature in narrow channels 
of superfluid helium propagate with the fourth-sound velocity.
     It is seen from the system of equations (\ref{7})$-$(\ref{9}) that the temperature and superfluid velocity discontinuities are of the first order as to the pressure discontinuity 
$\triangle p$. In this point there is a difference between the case of fourth-sound shock waves and the case of first- and second-sound shock waves.  When the discontinuity surface moves with the velocity of  first-sound wave, temperature jumps as well as normal- and superfluid-velocity difference jumps   are of the second order as to the pressure discontinuities. While in case of shock-front movement with the velocity of  second sound the pressure jumps prove to be of a higher order as to the discontinuities of the temperature and normal- and superfluid velocity difference.

\section{Conclusion}
The formation of shock waves of  finite-amplitude fourth sound  is studied.  The velocity of the discontinuity in the first approximation 
is found to be the same as the velocity of  fourth sound. This is similar to the cases of  first-and second-sound shock waves. 
But unlike these cases in fourth-sound shock waves  the temperature and superfluid velocity jumps turn to be  of the first order as 
to the pressure jump, that is these jumps are of the same order of magnitude.


\begin{thebibliography}{99}
\bibitem{A} K.R.Atkins, "Third and fourth sound in liquid He II", Phys. Rev.  {\bf 113}(4), 962, 1959. 
\bibitem{K} I.M.Khalatnikov, {\it Theory of superfluidity} (Benjamin, New York, 1965).
\bibitem{D} N.E.Dyumin  and E.Ya.Rudavski , "Temperature dependence of the attenuation of the fourth sound in He II",
 Low Temperature Physics  {\bf 1}(4), 521, 1975 (in Russian). 
\bibitem{L} Landau L.D., E.M. Lifshitz E.M., {\it Fluid mechanics} (Butterworth-Heinemann Ltd, 2000). 
\end{thebibliography}
\end{document}